\documentclass[
showpacs,preprintnumbers,amsmath,amssymb,nofootinbib,
]{revtex4}

\usepackage{amssymb}
\usepackage{bm}
\usepackage{amsfonts}
\usepackage{latexsym}
\usepackage{amsmath}

\usepackage [english]{babel}

\usepackage{array}

\usepackage[dvips]{epsfig}

\usepackage{epic}
\usepackage{eepic}
\usepackage{graphicx,psfrag}

\newcommand{\la}{\lambda}
\newcommand{\lb}{\label}

\newcommand{\al}{\alpha}
\newcommand{\cL}{{\cal L}}
\newcommand{\cN}{{\cal N}}

\newcommand{\ka}{\kappa}
\newcommand{\pa}{\partial}

\newcommand{\Ga}{\Gamma}
\newcommand{\fr}{\frac}
\newcommand{\lf}{\left(}
\newcommand{\rh}{\right)}

\newcommand{\bw}{\begin{widetext}}
\newcommand{\ew}{\end{widetext}}
\newcommand{\m}{{\mu\nu}}
\newcommand{\be}{\begin{equation}}
\newcommand{\ee}{\end{equation}}
\newcommand{\bea}{\begin{eqnarray}}
\newcommand{\eea}{\end{eqnarray}}
\newcommand{\nn}{\nonumber}

\newcommand{\dx}{{\dot x}}

\newcommand{\ba}{\begin{align}}
\newcommand{\ea}{\end{align}}
\newcommand{\s}{\sqrt{-g}}

\begin{document}


\title{Lorentz-covariant perturbation theory for relativistic
gravitational bremsstrahlung}

\author{Dmitri V. Gal'tsov, Yuri V. Grats and Alexander A. Matiukhin}
 \email{galtsov@physics.msu.ru}
\affiliation{Department of Theoretical Physics, Moscow State
University, 119899, Moscow, Russia}

\begin{abstract}
We  formulate Lorentz-covariant classical perturbation theory to
deal with relativistic bremsstrahlung under gravitational
scattering\footnote{This  is English translation of the Moscow State
University preprint issued in Russian in 1980 \cite{GGM0}. The
authors are grateful to Pavel Spirin for providing an electronic
version for Fig. 1. This submission was supported by the RFBR grant
08-02-01398-a.}. Our approach is a version of the fast motion
approximation scheme, the main novelty being the use of the momentum
space representation. Using it we calculate in a closed form the
spectrum of scalar, electromagnetic and gravitational radiation. Our
results for the total emitted energy agree with those by Thorne and
Kovacs. We also explain why the method of equivalent gravitons fails
to produce the correct result for the spectral-angular distribution
of emitted radiation under gravitational scattering, contrary to the
case of Weizs\"acker-Williams approximation in quantum
electrodynamics.
\end{abstract}


\maketitle

\section{Introduction}

Gravitational radiation by non-relativistic and quasi-relativistic
systems is low-multipole and can be easily calculated using the
quadrupole formula of General Relativity with higher multipole
corrections. With increasing velocities, the contribution from
higher multipoles becomes dominant, so one needs another technique.
The most adequate approach is the method of post-linear expansions
which was discussed in early 60-ies most notably by Bertotti
\cite{B}, Bertotti and Plebanski \cite{BP}, Havas and Goldberg
\cite{HG,Goldberg} (see also \cite{H,R,K,SH,In,An,Ros,CS,WeGo}). We
have developed a momentum space  version  of this approach
\cite{GG1} which is applied here to gravitational bremsstrahlung.
Although technically different, our calculations essentially overlap
and agree with those by Thorne and collaborators \cite{CT,KT3,KT4}.
The results also agree with an alternative calculations by Peters
\cite{Pe1,Pe2,Pe3} based on the linear perturbation theory on
Schwarzschild background. They disagree, however, with  calculations
based on ``equivalent gravitons'' method \cite{MaNu}, and we explain
the origin of this disagreement.

Other approaches to relativistic bremsstrahlung problem  are worth
to be mentioned. One, suggested by D'Eath \cite{DE}, is based on
replacing the boosted Schwarzschild metric by the impulsive
gravitational wave. Another, due to Smarr \cite{Sm1,Sm2}, appeals
to calculation of the radiation amplitude in the low-frequency
region. This overlaps with quantum calculation of the
cross-section in the Born approximation \cite{Barker}.

To calculate the leading order gravitational radiation in
relativistic collisions of particles interacting predominantly
through the non-gravitational forces it is enough to use the
linearized gravity on Minkowski background. In the case of
gravitational interaction we need at least the next post-linear
order. If one interprets the second order gravitational potentials
in terms of Minkowski space coordinates, one finds that the source
of gravitational radiation becomes non-local due to contribution of
the first order gravitational stresses (similarly for
non-gravitational radiation from particles interacting by gravity).
This non-locality leads to destructive interference of high
frequency part of radiation, so the spectrum will be different from
that of the electromagnetic bremsstrahlung.

\section{Field equations in quasilinear form}
Consider a system of point particles, interacting by
non-gravitational fields (scalar $\psi$ or massless vector
$A^\mu$) and moving in a self-consistent gravitational field
described by the metric $g_\m$. The action can be presented as  $
S=S_p+S_\psi + S_A+ S_g,$ where $S_p$ is the sum of particle
actions including non-gravitational interaction terms

\be \lb{1} S_p=-\sum\int(m+f\psi +eA_\mu \dx^\mu)ds, \ee $S_\psi$and
$S_A$ are scalar and Maxwell field actions \be\lb{2} S_\psi =\fr 1
{8\pi}\int
\partial_{\mu}\psi\,
\partial^{\mu}\psi\s d^4x,\quad  S_A =-\frac{1}{16\pi} \int F_\m F^\m \s d^4x,
\ee and the gravitational lagrangian is taken in the two-gamma form:
\be\lb{3} S_g=\int \cL \s d^4x,\quad \cL=-\fr1{2\ka^2}\int g^\m\lf
\Ga_{\mu\beta}^\alpha\Ga_{\nu\al}^\beta-\Ga_{\m}^\alpha\Ga_{\al\beta}^\beta\rh
 ,\quad \ka^2=8\pi G.\ee

Assuming gravitational field to be negligible at spatial infinity,
we choose asymptotically Minkowskian metric $\eta_\m$ in this region
and introduce the (non-tensor) metric deviation variable \be\lb{3a}
h_{\m }=g_{\m }-\eta _{\m }\ . \ee Then for $r \rightarrow \infty$
one has $h_{\m } \rightarrow 0$, but $h_{\m }$ are not necessarily
small everywhere. By convention, the indices of the quantities
$h_{\m },\
\partial _{\mu }$ and $\eta _{\m }$ will be raised and lowered by the Minkowski metric
$\eta _{\m }$ , while the indices of the true tensors are operated
with the metric $g_{\m }$.

Introducing an antisymmetric tensor density \be\lb{5}
H^{\al\nu\beta\la}=g\lf
g^{\al\la}g^{\beta\nu}-g^{\al\beta}g^{\la\nu}\rh,\ee one can present
 Einstein equations in a divergence form \be\lb{6} \lf
H^{\al\nu\beta\la}_{,\beta}\;g_{\la\mu}\;/\s\rh_{,\al}=-2\ka^2\s\lf
T^\nu_\mu+t^{\;\,\nu}_\mu\rh, \ee where $T^\nu_\mu$ is the total
matter stress-tensor, and $t^{\;\,\nu}_\mu$ is Einstein's canonical
pseudotensor \be\lb{7}
t^{\;\,\nu}_\mu\;=\;\fr1{\s}\;g^{\al\beta}_\mu \;\fr{\pa
(\s\;\cL)}{\pa g^{\al\beta}_\nu}-\delta^\nu_\mu \cL.\ee

Maxwell equations can be written in a similar form \be\lb{8} \lf
H^{\al\nu\beta\la}_{,\beta}\;g_{\la\mu}\;/\s\rh_{,\al}=-4\pi\;\s\;j^\mu,\ee
where the vector-current is \be\lb{9} j^\mu= \sum e\int
ds\;\dx^\mu\;\delta^4(x-x(s))/\s.\ee Finally, the scalar wave
equation reads \be\lb{10} \lf\psi_{,\mu}\;g^\m\;\s\rh_{,\nu} =4\pi
\;\s\; \tau,\ee with the scalar current \be\lb{11} \tau = \sum f\int
ds\; \delta^4(x-x(s))/\s.\ee Particle equations of motion
generically read \be\lb{12} \fr{d}{ds}\left[(m-f\psi)\right]\dx_\mu
=\fr{m}2 g_{\al\beta,\mu}\dx^\al\dx^\beta\;-mf \psi_{,\mu}\;+\;e
F_\m\;\dx^\nu.\ee

All the above equations are exact and  can be regarded as a system
defining the particle motion and evolution of the scalar,  vector
and gravity fields and  in a self-consistent way. However, since the
notion of delta-functions is not defined in the full non-linear
general relativity, we can deal with point particles only
perturbatively, expanding all the quantities in formal series in the
gravitational coupling $\ka$. For this one has to pass first to
quasilinear form of the field equations. For Einstein equations we
single out the linear part of the $H$-tensor: \be\lb{13}
H^{\al\nu\beta\la}=\eta^{\al\beta}\eta^{\la\nu}
-\eta^{\al\la}\eta^{\beta\nu}+\cL^{\al\nu\beta\la}+\cN^{\al\nu\beta\la}\ee
where $\cL^{\al\nu\beta\la}$ joins terms linear in $h_\m$:
\be\lb{14}\cL^{\al\nu\beta\la}=\psi^{\al\la}\eta^{\beta\nu}+\psi^{\beta\nu}\eta^{\al\la}-
\psi^{\al\beta}\eta^{\la\nu}-\psi^{\la\nu}\eta^{\al\beta},\quad
\psi^\m=h^\m-\fr12 \eta^\m\psi_\la^\la,\ee while
$\cN^{\al\nu\beta\la}$ denotes  non-linear in $h_\m $ terms.
Gravitational equations can be now presented as \be\lb{15}
\cL^{\la\al\beta\tau}_{,\al\beta}\eta_{\mu\la}\eta_{\nu\tau}
=2\ka^2\tau_\m,\quad \tau_\m=T_\m+S_\m,\ee where in $S_\m$ all the
non-linear terms are collected. To calculate gravitational radiation
one needs only  terms quadratic in $h_\m$.

Maxwell equations can be rewritten similarly: \be\lb{16}
A_{\al,\beta\nu}(\eta^{\al\mu}\eta^{\beta\nu}-\eta^{\al\nu}\eta^{\beta\mu})=
4\pi(j^\mu+S^\mu)\;\s, \ee where an effective ``gravitational''
vector current is given by \be\lb{17} \s S^\mu=\left[\lf 1/\s-1\rh
(\eta^{\al\mu}\eta^{\beta\nu}-\eta^{\al\nu}\eta^{\beta\mu})+ \lf
\cL^{\m\al\beta}+\cN^{\m\al\beta}\rh A_{\beta,\al}\right]_{,\nu}.\ee
In the scalar case we obtain similarly: \be\lb{18} \psi _{,\mu
}^{,\mu }=-4\pi f (\tau +S)\ , \ee where
$$ S=(1/{4\pi f})\sigma _{,\mu }^{\mu }\ ,\quad \sigma ^{\mu }=(\s g^{\m }-
\eta ^{\m })\psi _{,\nu }\ .$$ This looks  as the flat space wave
equation for the spin zero field, with an important difference,
however, that the ``source'' depends explicitly on $\psi$.

Now we can further simplify the quasilinear equations (which are
still exact in all orders in $\ka$) imposing the gauge conditions
\be\lb{19} \psi_{\m,\la}\eta^{\nu\la}=0,\qquad A_{\mu,\nu}\eta^\m=0,
\ee which are consistent with the field equations by virtue of the
identities: \be\lb{20} \tau_{\m,\la}\eta^{\nu\la}=0,\qquad \lf\s\;
S^\mu \rh_{,\mu}=\lf\s\; j^\mu \rh_{,\mu}.\ee In this gauge Einstein
and Maxwell equations read \be\lb{21} \Box \psi_\m=2\ka^2
\tau_\m,\quad \Box A_\mu=-4\pi \s (j^\la+S^\la)\eta_{\la\mu},\ee
with $\Box =-\pa_{\la}\pa_{\tau}\eta^{\la\tau}$.

\section{Scalar bremsstrahlung under gravitational scattering }
Consider two point masses $m_1$ and $m_2$, one of which   ($m_1$)
carries the scalar charge $f$. Particles interact via gravity and
the systems emits both gravitational and  scalar waves. In this
section we calculate scalar radiation. The action reads \be\lb{22}
  S=-\int (m_1+f \psi(x) )\sqrt{\dot{x_1}^2} ds
-m_2\int \sqrt{\dot{x_2}^{2}} ds  +\frac{1}{8\pi}\int \s d^Dx
\partial_{\mu}\psi\,
\partial^{\mu}\psi\,
 \ee where
 $\dot{x}^2\equiv
g_{\alpha\beta}\dot{x}^\alpha\dot{x}^\beta$,  dot denotes
differentiation with respect to the interval $s$, and the metric
signature is mostly minus.

The full system of equations describing the collision consists of
Einstein equations, scalar field equation and particle equations as
given in the previous section.  The total loss of the four-momentum
during the collision can be presented as \be \lb{24}\Delta P_S^{\mu
}=\int _{-\infty }^{\infty }dt \oint T_{\psi }^{\mu i }d\sigma _i \
, \ee where \be\lb{25} T_{\psi }^{\m
}=\frac1{4\pi}\left(\psi_{,\mu}\psi_{,\nu}-\frac12
\psi_{,\alpha}\psi^{,\alpha}\right)\ee is the energy-momentum tensor
of the massless field $\psi $, and integration is performed over the
sphere of infinite radius. The non-zero contribution comes from the
terms in $T_{\psi }^{\mu i}$, which fall off at infinity as
$r^{-2}$. Also, without changing the integral, one can add to the
integrand the total derivative over time $(\psi ^{,\mu }\psi
^{,0}-(1/2)\eta ^{\mu 0}\psi _{,\alpha }\psi ^{,\alpha } )_{,0}$.
Then, applying the Gauss theorem we can transform the Eq. (\ref{24})
to the following form: \be\lb{26} \Delta P_S^{\mu }=\frac{1}{4\pi
}\int d^4x \ \psi ^{,\mu }\psi _{,\nu }^{,\nu }\ . \ee To exclude an
infinite self-energy part it is enough to substitute as  $\psi
_{,\nu }^{,\nu }$ the right hand side of the Eq. (\ref{18}), while
as $\psi $ -- the $t$-odd part of the retarded potential \be\lb{27}
\psi (x) =-\frac{if}{(2\pi )^2}\int d^4k e^{-ikx}\varepsilon (k^0)
\delta (k^2)[\tau (k)+ S(k)]\ , \ee where $\varepsilon (k^0)=\theta
(k^0)-\theta (-k^0)$ is the sign function and the Fourier transforms
are defined as \be\lb{28} \tau (k) =\int d^4x e^{ikx}\tau (x)\ \ee
and similarly for $S(k)$. We obtain: \be\lb{29} \Delta P_S^{\mu
}=\frac{f^2}{2\pi ^2}\int d^4k k^{\mu }\theta (k^0)\delta
(k^2)\mid\tau (k)+ S(k) \mid^2\ . \ee This expression is analogous
to the usual one in electrodynamics, differing from it by presence
of the non-local current $S(k)$ which we will call the stress
current.

To find $\tau (k)$ and $S(k)$ we will solve the Einstein equation,
the particle equations and the scalar field equation  (\ref{18})
expanding $g_{\m },\ \psi $ and $x^{\mu }$ in powers of the
gravitational constant $G$. The actual expansion parameter in the
ultrarelativistic collision problem will be the ratio of the
gravitational radius of one of the particle to the impact parameter.
One can show that approximation is valid if the particle scattering
angle is small with respect to the radiation angle \cite{We}
\be\lb{30} G(m_1+m_2)/{\rho v^2}\ll 1/{\gamma }\ , \ee where $\gamma
= (1-v^2)^{-1/2},\ v $ -- the relative velocity of the colliding
particles, $\rho $ -- the impact parameter.

We parameterize the particles world lines as  \begin{align}
\lb{31} & x_1^{\mu }= \Delta ^{\mu }+(p_1^{\mu }/m_1)s_1 +\tilde
x_1^{\mu }(s_1)\ , \nn \\  & x_2^{\mu }=(p_2^{\mu }/m_2)s_2 +
\tilde x_2^{\mu }(s_2)\ ,\end{align}  with $\sqrt{-\Delta ^2}=\rho
,(\Delta p_1)=(\Delta p_2)=0, \Delta ^{\mu }$ -- is the
four-vector which in the rest frame of the second particle takes
the form $(0, \vec\rho)$. Here and below we use brackets
$(\ldots)$ to denote scalar products with respect to Minkowski
metric. We choose the initial conditions \be\lb{32} \tilde
x_a^{\mu }(-\infty )= d\tilde x_a^{\mu }/ds(-\infty )=0,\ a=1,2\ ,
\ee so that   $p_a^{\mu }$ is the four-momentum of the particle
$a$ before the collision, and $\eta ^{\m }p_a^{\mu }p_a^{\nu
}=m_a^2; \ \tilde x_a^{\mu }$ is the correction due to the
gravitational interaction.

In the lowest order in gravitational interaction the correction to
the space-time metric $\eta _{\m }$ due to the second particle reads
 \be \lb{33}
h_2^{\m }(x)= \frac{2G}{\pi ^2}\left(p_2^{\mu }p_2^{\nu
}-\frac{1}{2}m_2^2\eta ^{\m }\right)\int d^4k \frac{\delta
(kp_2)}{k^2}e^{-ikx}\ . \ee Substituting this into the equation of
motion of the particle $m_1$ we find \be\lb{34} \tilde x_1^{\mu
}(s) = -i\frac{G}{\pi ^2}\int d^4 q \frac{\delta (qp_2)}
{q^2(qp_1)^2}e^{-iq(\Delta
+\frac{p_1}{m_1}s)}\biggl\{2(qp_1)\left[(p_1p_2)p_2^{\mu
}-\frac{m_2^2}{2}p_1^{\mu }\right]-
\left[(p_1p_2)^2-\frac{(m_1m_2)^2}{2}\right]q^{\mu }\biggr\}\ .\ee
To calculate the Fourier-transform of $S$ we use the expression
for $\psi $ in the lowest order (zero order in $G$) \be\lb{35}
\psi (x) =\frac{fm_1^2}{2\pi ^2}\int d^4k \frac{\delta
(kp_1)}{k^2} e^{-ik(x-\Delta )}\ . \ee The Fourier-transforms
$\tau (k)$ and $S(k)$ can be computed as follows. Using the
integrals
\begin{align} \lb{36} & \int d^4q e^{-iq\Delta }\frac{\delta (qp_2)\delta (qp_1 -kp_1)}{q^2}=
-\frac{2\pi }{I}K_0(z_1)\ ,\nn \\ & \int d^4q e^{-iq\Delta }q^{\mu
}\frac{\delta (qp_2)\delta (qp_1-kp_1)}{q^2}= 2\pi
\frac{m_2^2(kp_1)}{I^3} \biggl\{ \biggl[ p_1^{\mu
}-\frac{(p_1p_2)}{m_2^2}p_2^{\mu }\biggr] K_0(z_1)-i(kp_1)\Delta
^{\mu }\frac{K_1(z_1)}{z_1}\biggr\}\ ,
\end{align} we obtain
\begin{align}  \lb{37} \tau (k)=-4Gm_1m_2 e^{ik\Delta }& \biggl\{ \biggl[
\biggl( 1-\frac{(m_1m_2)^2} {2I^2}\biggr)
\frac{(p_1p_2)}{I}\frac{z_2}{z_1}+\frac{(m_1m_2)^3}{2I^3} \biggr]
K_0(z_1)- \nn \\ &-i(k\Delta )\frac{m_1m_2}{I}\biggl(
1+\frac{(m_1m_2)^2}{2I^2}\biggr) \frac{K_1(z_1)}{z_1}\biggr\}\
,\end{align} where $I=\{(p_1p_2)^2-(m_1m_2)^2\}^{1/2}\ ,\quad
z_1=\sqrt{-\Delta ^2}(kp_1)m_2 I^{-1}\ ,\quad z_2=\sqrt{-\Delta
^2}(kp_2)m_1I^{-1}\,\quad  K_{0,1}$ are the Macdonald functions.
Lorentz-invariant integrals  (\ref{36}) an be conveniently
computed in the rest frame of the second particle  $p_2^{\mu
}=(m_2,0)$. Then two integrations from four are performed using
the delta-functions, while the remaining two-dimensional integral
in the plane orthogonal to ${\bf p_1}$ is computed using the polar
coordinates. The angular integral is standard, and the last one is
done by contour integration.

Integration over $d^4q$ in the expression for $S(k) $ can be done
using the Feynman parameterization. We obtain \be\lb{38}
S(k)=4GIe^{ik\Delta }z_2^2\int _0^1 dx e^{-ix(k\Delta
)}\frac{K_1(z(x))} {z(x)}\ , \ee where
$$z(x)=\sqrt{\xi ^2(x)};\ \xi^{\mu }(x)=(1-x)z_1p_2^{\mu
}/m_2+xz_2p_1^{\mu }/ m_1\ ,$$
$$ z(0)=z_1=\sqrt{-\Delta ^2}(kp_1)m_2/I;\ z(1)=z_2=\sqrt{-\Delta ^2}
(kp_2)m_1/I\ .$$ The Eqs. (\ref{37}) and  (\ref{38}) are obtained
under the only restriction  (\ref{30}), they are valid for arbitrary
velocities $v$. In the rest frame of the second particle $p_2^{\mu
}=(m_2, 0)$ we further specify the coordinates so that
$$ \vec\rho=(0 ,0, \rho)\ ,\qquad
\vec p_1=(0 , p_1, 0)\ ,$$
$$ \vec k=\omega(\sin\theta \sin\phi, \cos\theta, \sin\theta \cos\phi)\ .$$
Consider the case of non-relativistic velocities. For $v\ll 1, \,
z(x)=z_1=z_2=\omega \rho /v$, then the integral in (\ref{38}) is
easily done and we obtain \be\lb{39} \tau
(k)=\frac{2Gm_1m_2}{v^2}e^{-i\omega \rho \sin \theta \cos \phi }
[\cos \theta K_0(a)-i\sin \theta \cos \phi K_1^2(a)]\ , \ee
\be\lb{40} S(k)=4Gm_1m_2vaK_1(a)e^{-i\omega \rho \sin \theta \cos
\phi }\ , \ee where $a=\omega \rho /v$. From (\ref{39}) and
(\ref{40})  it is clear that $S/\tau \sim v^3$, so in the lowest in
$v$ approximation radiation is entirely determined by the local
current (\ref{39}). Substituting it into the Eq. (\ref{29}), after
some simple transformations we find \be\lb{41} \frac{d^2E_S}{d\omega
d\Omega }=(\frac{Gfm_1m_2}{\pi \rho v})^2a^2 [\cos ^2 \theta
K_0^2(a)+\sin ^2\theta \cos ^2\phi K_1^2(a)]\ . \ee One can see that
for small velocities the characteristic radiation frequency $\omega
\sim v/\rho $ is inverse to the effective time of collision $\rho
/v$.

The total energy loss during the collision can be obtained
integrating (\ref{41}) over angles and the  frequency: \be\lb{42}
\Delta E_S=(\pi /6)(fGm_1m_2)^2/\rho ^3v\ . \ee In the
ultrarelativistic case $(\gamma \gg 1)$ the effective spread of the
stress current $S(x)$ is of the order of $\rho $. So it can be
expected that for the wavelengthes   $\lambda \gg \rho $ the source
with act as point-like. Indeed, for $\omega \ll \rho ^{-1} \
(\lambda \gg \rho )$ the argument of the Macdonald functions
(\ref{37}) and (\ref{38}) is small for all values of parameters, and
with account for the leading terms we obtain \be\lb{43} \tau
(k)=-i\frac{4Gm_1m_2}{\gamma }\frac{\sin \theta \cos \phi }{\omega
\rho
 \delta ^2 },\ \frac{S(k)}{\tau (k)}\sim \omega \rho \ll 1\ ,
\ee where $\delta =1-v\cos \theta $. Substituting (\ref{43}) into
(\ref{29}) we find: \be\lb{44} \frac{dE_S}{d\omega
}=\frac{16}{3\pi }\frac{(fGm_1m_2)^2}{\rho ^2}\gamma ^2,\ \omega
\ll \rho ^{-1}\ . \ee In the frequency region $\omega \geqslant
\rho ^{-1}$ the contributions of the local and the non-local
currents are of the same order. In this case for the spectral
distribution of the total emitted energy we obtain: \be\lb{45}
\frac{dE_S}{d\omega }=\frac{16(fGm_1m_2)^2}{\pi }\omega ^2
\int_0^{\infty }\int_0^{\infty} d\xi d\eta \frac{e^{-\frac{2\omega
\rho }{\gamma } \sqrt{1+\xi ^2}\sqrt{1+\eta ^2}}}{(1+\xi
^2)^{3/2}(1+\eta ^2)^{1/2}} \ln \frac{1+\xi ^2+ \eta ^2}{\eta ^2}\
.\ee For  $ \rho ^{-1}\le \omega \ll \gamma /\rho $ \be\lb{46}
\frac{dE_S}{d\omega }=\lambda _S\frac{(fGm_1m_2)^2}{\rho ^2}\gamma
^2\biggl( \frac{\omega \rho }{\gamma }\biggr)^2\ , \ee where
$$ \lambda _S =\frac{64}{3\pi }\int _0^{\infty } dxx^{-3}\ln ^3(x+\sqrt{1+x^2})
\approx 8.$$ For relatively high frequencies  $\omega \gg \gamma
/\rho $ the integral in (\ref{45}) is formed at  $\xi ,\ \eta \ll
1$. So approximately \be\lb{47} \frac{dE_S}{d\omega }\simeq
\frac{4(fGm_1m_2)^2}{\rho ^2}\gamma ^2 \biggl( \frac{\omega \rho
}{\gamma }\biggr) \ln \frac{4e^C\omega \rho }{\gamma }
e^{-\frac{2\omega \rho }{\gamma }}\ . \ee

The expressions (\ref{44}), (\ref{46}) and (\ref{47}) together
describe the behavior of the spectral curve. It follows, in
particular, that in the spectral distribution there is a maximum
around the frequency  $\omega \sim \gamma /\rho $. The total energy
loss during the collision is obtained integrating
 (\ref{45}) over the frequency:
 \be\lb{48} \Delta E_S=\Lambda_S\frac{(fGm_1m_2)^2}{\rho
^3}\gamma ^3,\ \Lambda_S=\frac{ 3\tilde G}{2}+\frac{77}{12}-2\pi
\simeq 1.51\ , \ee where $\tilde G$ is the Catalan constant.

Let us compare these results with the case of the electromagnetic
interaction in Minkowski space, when the source term in the equation
for $\psi $ does not contain the stress current $S(x)$. Suppose that
the colliding particles are electrically charged
 $(e_1,e_2)$ and neglect gravitational interaction with respect to electromagnetic one.
Then as the source  $\tau (k)$ in  (\ref{29}) one has to use the
Fourier-transform of the trace of the particles energy-momentum
tensor. After similar calculations we obtain: \be \lb{49}
T(k)=\frac{2e_1e_2(m_1m_2)^2}{I^3}e^{ik\Delta }\biggl\{ \biggl[
(p_1p_2)-m_1^2\frac{(kp_2)}{(kp_1)}\biggr] K_0(z_1) -i(k\Delta
)(p_1p_2) \frac{K_1(z_1)}{z_1}\biggr\}\ . \ee In the
ultrarelativistic case $(\gamma \gg 1)$ the spectral-angular
distribution is dominated by the second term in (\ref{49}). In the
rest frame of the second particle we find in the leading order in
$\gamma $: \be\lb{50} \frac{dE_S}{d\omega }=\frac{8}{\pi
}\frac{(fe_1e_2)^2}{\rho ^2}z \int _z^{\infty
}dx\frac{z}{x}\biggl( 1-\frac{z}{x} \biggr) K_1^2(x)\ , \ee where
$z=\omega \rho /2\gamma ^2$. Using the asymptotic expansions for
the Macdonald functions for small and large arguments, from
(\ref{50}) we find for $\omega \ll \gamma ^2/\rho $: \be\lb{51}
\frac{dE_S}{d\omega} =\frac{4}{3\pi}\frac{(fe_1e_2)^2}{\rho ^2}\ ,
\ee while for high frequencies $\omega \gg \gamma ^2/\rho $
\be\lb{52} \frac{dE_S}{d\omega }=2\frac{(fe_1e_2)^2}{\rho
^2}\frac{\gamma ^2}{\omega \rho } e^{-\frac{\omega \rho }{\gamma
^2}}\ . \ee

Note, that for the local source case our methods gives the energy
loss without restrictions on the relative velocity of collision.
Indeed, substituting  (\ref{49}) into  (\ref{29}) and integrating
over frequencies and angles we obtain \be\lb{53} \Delta E_S=
\frac{\pi }{8}\frac{(fe_1e_2)^2}{v\rho ^3}\biggl( \gamma ^2+
\frac{1}{3} \biggr)\ . \ee Thus we see that there is substantial
difference between the spectrum of the bremsstrahlung from
gravitational scattering and that in the case of electromagnetic
interaction.  In the first case there is a maximum at  $\omega
\sim \gamma /\rho $, while the spectral distribution (\ref{50}) is
monotonous function of the frequency. For gravitational
interaction the exponential cut off corresponds to the frequency
$\omega \geqslant \gamma /\rho$, and not to $\omega \geqslant
\gamma ^2/\rho$ as in the case of the electromagnetic scattering.
Finally, the total energy loss at gravitational scattering
(\ref{48}) is $\gamma $ times less that the corresponding quantity
in the electromagnetic case  (\ref{53}) for the same scattering
angle, i.e. under the condition   $Gm_2m_1\gamma \sim e_1e_2 $.

These properties can be qualitatively explained by the presence of
the non-local (in terms of the flat space-time picture)
stress-current source in the equation for the radiated field  $\psi
$ in the case of gravitational scattering. This current has and
effective transverse dimension of the order of $\rho $ and
longitudinal of the order of $\rho /\gamma $ ( $\gamma $ times
smaller  due to the Lorentz contraction). For large wavelengthes
($\lambda \gg \rho $) the source non-locality is insignificant and
the low frequency limit is the same as for the electromagnetic
interaction case, when there is no non-local term at all. For
$\lambda \le \rho $ radiation from the most distant elements of the
source exhibit a destructive interference for the angles close to
$\pi /2$, which leads to the  gap in the spectrum. Finally, for
$\lambda < \rho /\gamma $ the conditions for destructive
interference are fulfilled for the forward direction, in which the
most of the energy is emitted. This leads to substantial decrease of
the radiation.

\section{ Electromagnetic bremsstrahlung under gravitational scattering
} The case of the electromagnetic interaction is rather similar. Let
the particle  $m_1$  carry the electric charge  $e_1$. Using
analysis of the Sec.2 we can present Maxwell equations as follows:
\be\lb{54} (\eta ^{\mu \alpha }\eta ^{\nu \beta }F_{\alpha \beta
})_{,\nu }=-4\pi (J^{\mu } +S^{\mu })\ , \ee where the
stress-current is \be\lb{55}
 S^{\mu }=\sigma^{\mu \nu }_{,\nu }\ ,
\ee
$$ \sigma ^{\mu \nu } =(1/4\pi )(\s g^{\mu \alpha }g^{\nu \beta }-
\eta ^{\mu \alpha }\eta ^{\nu \beta })F_{\alpha \beta }\ ,$$
\be\lb{56} J^{\mu }= e_1 \int ds \frac{dx_1^{\mu }}{ds} \delta
(x-x_1(s))\ . \ee Imposing the flat space Lorentz gauge on  the
four-potential $A_{\mu }$: \be\lb{57} \eta ^{\mu \nu } A_{\mu ,\nu
}=0\ , \ee we cast Maxwell equations into the form convenient for
iterative solution: \be\lb{58} \eta ^{\mu \nu }\eta ^{\alpha \beta
}A_{\nu , \alpha ,\beta }=4\pi (J^{\mu }+S^{\mu })\ . \ee

It is convenient to choose two linearly independent polarization
vectors as \be\lb{59} e^{\mu }_{\phi }=\lambda _{\phi }e^{\mu \nu
\rho \sigma }k_{\nu }p_{1\rho } p_{2\sigma }\ , \quad e^{\mu
}_{\theta }=\lambda _{\theta }e^{\mu \nu \rho \sigma }k_{\nu
}e_{\phi \rho } p_{2\sigma }\ , \ee \be\lb{60} \lambda _{\phi }=
(-{\it P}^2)^{-1/2}\ , \quad {\it P}^{\mu }=(kp_2)p_1^{\mu }-
(kp_1)p_2^{\mu }\ ,\quad \lambda _{\theta}=(kp_2)^{-1}\ . \ee They
satisfy the following conditions: \be\lb{61} (e_{\theta }e_{\phi
})=(ke_{\theta })=(ke_{\phi })=0\ ,\quad (e_{\phi
}e_{\phi})=(e_{\theta }e_{\theta})=-1 \ee and in the rest frame of
the second particle read: $e_{\theta }^{\mu } =(0, \vec e_{\theta
}),\ e_{\phi }^{\mu }=(0, \vec e_{\phi})\ ,$ where $\vec e_{\theta}$
and $\vec e_{\phi }$ are unit vectors along $\theta$ and $\phi$.

The expression for the four-momentum loss due to electromagnetic
interaction with polarization $\lambda \ (\lambda =\theta ,\ \phi
)$ can be derived analogously to  the Eq. (\ref{29}) and reads:
\be\lb{62} \Delta P_{em}^{(\lambda)\mu } = \frac{1}{2\pi ^2}\int
d^4k k^{\mu }\theta (k^0) \delta (k^2)\mid I^{(\lambda)}(k)\mid
^2\ , \ee where $ I^{(\lambda)}(k)=\eta _{\alpha \beta }e_{\lambda
}^{\alpha }(J^{\beta }(k) + S^{\beta }(k))$. As in the scalar
case, one has to retain in  $S^{\mu }$ only terms falling off
asymptotically as $r^{-2}$. In this approximation \be\lb{63}
S^{\mu }(x)=-(1/4\pi )(F_{\sigma }^{\mu }h^{\sigma \nu }+F_{\sigma
}^{\nu } h^{\mu \sigma }-(h_{\sigma }^{\sigma }/2)F^{\mu \nu
})_{,\nu }\ . \ee The subsequent calculations are similar to the
scalar case. The Fourier-transforms of the currents  (\ref{56})
and (\ref{63}) are computed in the full analogy with the previous
section resulting in
\begin{align} \lb{64} J^{\mu }(k)=&
\frac{4G}{(kp_1)}e^{ik\Delta }\biggl\{\frac{(p_1p_2)}{I}\biggl(
1-\frac{(m_1m_2)^2}{2I^2}\biggr)K_0(z_1)[(kp_1)p_2^{\mu
}-(kp_2)p_1^{\mu }]+ \nonumber \\ &+i\frac{m_2}{\sqrt{-\Delta
^2}}\biggl(1+\frac{(m_1m_2)^2}{2l^2}\biggr)K_1(z_1) [(k\Delta
)p_1^{\mu }- (kp_1)\Delta ^{\mu }]\biggr\}\ ,
\end{align}
\begin{align}  S^{\mu}(k) = &- \frac{4Gm_2^2}{I}e^{ik\Delta }\, \int _0^1 dx e^{-ix(k\Delta )}
\biggl\{-\Delta
^2\biggl[\frac{(kp_2)}{m_2^2}-x\frac{(kp_2)}{m_2^2}\biggl(
1+\frac{(m_1m_2)^2}{2I^2}\biggr)- \nonumber \\
&-(1-x)\frac{(kp_1)(p_1p_2)}{2I^2}\biggr] \frac{K_1(z(x))}{z(x)}
\biggl( (kp_1)p_2^{\mu }-(kp_2)p_1^{\mu }
\biggr)-i\frac{(p_1p_2)}{m_2^2}K_0(z(x)) \times \nonumber \\
\lb{65} & \hphantom{(1-x)(1-)} \times \biggl( (kp_2)\Delta ^{\mu
}- (k\Delta )p_2^{\mu }\biggr) - \frac{1}{2} K_0(z(x))\biggl(
(k\Delta )p_1^{\mu }- (kp_1)\Delta ^{\mu }\biggr) \biggr\}\ .
\end{align} In (\ref{65}) terms, proportional to $k^{\mu }$ are
omitted since they do not contribute to radiation by virtue of
(\ref{61}).

For small relative velocity  $(v\ll 1)$ radiation is generated
predominantly by the local current $J^{\mu }$, since $S^{\lambda
}/J^{\lambda }\sim v^2$. In this case
$$\frac{d^2E_{em}^{\theta }}{d\omega d\Omega }=\biggl( \frac{e_1Gm_2}{\pi \rho v }
\biggr) ^2\sin ^2\theta \ a^2[K_0^2(a)+ctg ^2\theta \sin ^2\phi
K_1^2(a)]\ ,$$ \be\lb{66} \frac{d^2E_{em}^{\phi }}{d\omega d\Omega
}=\biggl( \frac{e_1Gm_2}{\pi \rho v} \biggr) ^2 \cos ^2 \phi a^2
K_1^2(a)\ . \ee Integrating over frequencies and angles we get
$$\Delta E_{em }^{\theta }=\frac{7\pi }{48}\frac{(e_1Gm_2)^2}{v\rho ^3}\ ,$$
\be\lb{67} \Delta E_{em}^{\phi }=\frac{3\pi
}{16}\frac{(e_1Gm_2)^2}{v\rho ^3}\ . \ee

For ultrarelativistic collisions  $(\gamma \gg 1)$ in the low
frequency range $(\omega \ll 1/\rho )$ contribution of the non-local
stress current is relatively small, $S^{\lambda }/J^{\lambda }\sim
\omega \rho \ll 1$, and we have:
$$\frac{dE_{em}^{\theta }}{d\omega} =\frac{8}{3\pi}\frac{(e_1 Gm_2\gamma)^2}
{\rho ^2}\ ,$$ \be\lb{68} \frac{dE_{em}^{\phi }}{d\omega}
=\frac{8}{\pi}\frac{(e_1 Gm_2\gamma)^2} {\rho ^2}\ . \ee For
$\omega \geqslant \rho ^{-1}$ in the leading in  $\gamma $
approximation the spectral distribution of the radiated energy is
given by
$$
\frac{dE_{em}}{d\omega }=\frac{16}{\pi }(e_1Gm_2)^2\omega ^2\int
_0^{\infty } \int _0^{\infty }d\xi d\eta e^{-\frac{2\omega \rho
}{\gamma }\sqrt{1+\xi ^2} \sqrt{1+\eta ^2}} \times $$ \be\lb{69}
 \times \frac{1+2\xi ^2}{(1+\xi ^2)^{3/2}(1+\eta ^2)^{1/2}}\ln \frac{
1+\xi ^2 +\eta ^2}{\eta ^2}\ . \ee In (\ref{69}) we performed
summation over polarizations.

Using (\ref{69}) one can show that for $\rho ^{-1} \le \omega \ll
\gamma /\rho $ the spectral distribution behaves as follows:
\be\lb{70} \frac{dE_{em}}{d\omega }=(e_1Gm_2)^2\biggl( \frac{\gamma
}{\rho }\biggr) ^2 \biggl( \frac{\omega \rho }{\gamma }\biggr) \ln
\frac{\gamma }{\omega \rho }\ , \ee while for the frequencies
$\omega \gg \gamma /\rho $ \be\lb{71} \frac{dE_{em}}{d\omega
}=4(e_1Gm_2)^2\biggl( \frac{\gamma }{\rho }\biggr) ^2 \biggl(
\frac{\omega \rho }{\gamma }\biggr) \ln \frac{4e^C\omega \rho
}{\gamma } e^{-\frac{2\omega \rho }{\gamma }}\ . \ee Comparing
(\ref{68}), (\ref{70}) and (\ref{71}) one can notice the fall off in
the spectrum in the frequency range $\omega \sim \rho ^{-1} $ and
the maximum at $\omega \sim \gamma /\rho $ (Fig. 1).

For the total energy loss we obtain: \be\lb{72} \Delta
E_{em}=\Lambda _{em}\frac{(e_1Gm_2)^2\gamma^3}{\rho ^3}\ , \ee where
$\Lambda _{em}=5\tilde G/2 +43/12 -\pi \approx 2.75$. Splitting on
polarizations is given by  $\Lambda _{em }\rightarrow
\Lambda_{em}^{\lambda }\ (\Lambda _{em}^{\theta }\approx 1.75,\
\Lambda _{em}^{\phi }\approx 1.00).$ Our result (\ref{72})
qualitatively agrees with that of \cite{Pe1} but differs from that
of \cite{MaNu} by  absence of the factor $\ln 2\gamma $.

In the case of both particles electrically charged with large
charge to mass ratio in geometric units one can neglect
gravitational interaction and the bremsstrahlung problem is
simplified considerably. Then the stress-current   $ S^{\mu }=0$,
and the radiation amplitude is fully given by the local current.
Consider for simplicity the case $m_2\gg m_1$. Then the
Fourier-transform of the current is given by \be\lb{73} J^{\mu
}(k)=\frac{2(e_1e_2)(m_1m_2)^2}{I^3}e^{ik\Delta }\biggl\{ \biggl(
p_2^{\mu }-\frac{(kp_2)}{(kp_1)}p_1^{\mu }\biggr) K_0(z_1)
 +i\frac{(p_1p_2)}{m_1^2}((kp_1)\Delta ^{\mu }-(k\Delta
)p_1^{\mu })\frac{ K_1(z_1)}{z_1}\biggr\}\ . \ee
In the
non-relativistic case
 $(v\ll 1)$ the spectral-angular distribution of the emitted energy, as it can be expected,
is given by the Eq. (\ref{66}) with the substitution  $Gm_1m_2
\rightarrow e_1e_2$. As before, two independent polarization states
are given by the unit vectors  (\ref{59}), (\ref{60}). Using the
Eqs.  (\ref{73}), (\ref{59}) and (\ref{60})  and passing to the rest
frame of $m_2$ one finds with account for  (\ref{62}) the following
expression for the energy loss due to electromagnetic radiation with
the polarization  $\lambda \ (\lambda =\theta ,\phi )$:
$$\Delta E_{em}^{\theta }=\frac{7\pi }{64}\frac{(e_1^2e_2)^2}{m_1^2\rho ^3v}
\biggl( \gamma ^2+\frac{1}{3}\biggr)\ ,$$ \be\lb{74} \Delta
E_{em}^{\phi }=\frac{9\pi }{64}\frac{(e_1^2e_2)^2}{m_1^2\rho ^3v}
\biggl( \gamma ^2+\frac{1}{3}\biggr)\ . \ee

Note that Eqs. (\ref{74}) are valid for an arbitrary relative
velocity of the colliding particles. In the ultrarelativistic case
$(\gamma \gg 1)$ the spectral distribution is given by the second
term in  (\ref{73}), so in the leading approximation in   $\gamma $
\be\lb{75} \frac{dE_{em}}{d\omega }=\frac{4}{\pi
}\frac{(e_1^2e_2)^2}{m_1^2\rho ^2}z \int _z^{\infty }dx \biggl(
1-\frac{2z}{x}+\frac{2z^2}{x^2}\biggr) K_1^2(x)\ , \ee where
$z=\omega \rho /2\gamma ^2$. In the low-frequency limit $\omega \ll
\gamma ^2/\rho $ the Eq. \ref{75} has the form \be\lb{76}
\frac{dE_{em}}{d\omega} =\frac{8}{3\pi}\frac{(e_1^2e_2)^2}{m_1^2\rho
^2}, \ee which coincides with (\ref{68}), if both results are
expressed in terms of the scattering angle. At high  frequencies
$\omega \gg \gamma ^2/\rho $ the spectral distribution has
exponential cut off: \be\lb{77} \frac{dE_{em}}{d\omega}
=\frac{(e_1^2e_2)^2}{m_1^2\rho ^2} e^{-\frac{\omega \rho }{\gamma
^2}}\ . \ee The numerical curve for the spectral distribution is
given in Fig. 1.

Comparing the Eqs. (\ref{68}), (\ref{70}), (\ref{71}) and (\ref{72})
with the Eqs. (\ref{74}), (\ref{76}) and (\ref{77}) one can see that
the difference between spectral properties  of radiation for
particles interacting by gravity and by non-gravitational forces is
similar for scalar and electromagnetic radiation.

\section{Gravitational bremsstrahlung}
Consider now the system of two gravitating point particles $m_1$ and
$m_2$. We choose coordinates in such a way that the metric
perturbations be small at infinity when particles are at finite
distance from each other. Then we can treat the particles at $t=\pm
\infty $ as free and the metric to be flat (excluding the self-field
of each particle in its vicinity which can be removed by classical
renormalization). Denote the covariant components of the 4-momenta
as \be\lb{78} p_{\mu }^a=\lim _{s\rightarrow -\infty }m_au_{\mu }^a\
,\quad u_{\mu }^a=g_{\mu \nu } dx_a^{\nu }/ds\ ,\quad
{p'}_{\mu}^{a}=\lim _{s\rightarrow \infty }m_a{u'}_{\mu }^a\ , \quad
a=1,2\ .\ee The change of the total four-momentum of the system is
due to radiation friction acting on the particles. Although for two
relativistic gravitationally interacting particles it is problematic
to find the gauge independent local reaction force, one can still
find in a coordinate independent way the expression for the total
momentum loss during the whole collision time: \be\lb{79}  \Delta
P_{\mu }=\sum _{a=1,2} (p_{\mu }^{'a}-p_{\mu }^{a})=\sum
_{a=-1,2}m_a\int _{-\infty }^{\infty }ds\frac{du_{\mu }^a}{ds}. \ee
This quantity can be shown to be independent on the coordinate
choice if the coordinate transformation preserve the above
asymptotic conditions. Using the equations of motion we find
\be\lb{80}  \Delta P_{\mu }=\frac{1}{2}\int d^4x \s g_{\nu \sigma ,
\mu }T^{\nu \sigma }\ . \ee Since the covariant derivative of the
stress-tensor is zero $T_{\mu ;\nu }^{\nu }=0$, we have \be\lb{81}
g_{\nu \sigma ,\mu }\s T^{\nu \sigma }=2(\s T_{\mu }^{\nu })_{,\nu}\
. \ee Now we make use of the conservation equation
 \be
\left[\s \left( T_{\mu}^{\nu }+t_{\mu }^{\nu }\right)\right]_{,\nu
}=0,
 \ee where $t_{\mu }^{\nu }$ is the Einstein pseudotensor.
In our approximation it will be enough to keep only quadratic terms
in $h_{\mu\nu}$:
$$ t^{\m}=\frac{1}{32\pi G}\Bigl[{\psi_{\alpha\beta}}^{,\mu}(\psi^{\alpha\beta ,\nu}-
2{\psi_{\alpha\nu}}^{,\beta}-\frac{1}{2}\eta^{\alpha\beta}\psi^{\nu})-$$
$$-\eta^{\m}(\psi_{\alpha\beta ,\lambda}\psi^{\alpha\beta ,\lambda}-
2\psi_{\alpha\beta ,\lambda}\psi^{\alpha\lambda ,\beta}-\frac{1}{2}
\psi_{\lambda}\psi^{\lambda})\Bigr]\ .$$  As a result, we transform
the momentum loss to the form   \be\lb{82} \Delta P_{\mu }= -\int
d^4x (\s \ t_{\mu }^{\nu })_{,\nu }. \ee We assume the gauge
$\psi^{\mu\nu}_{,\nu}=0$ and calculate the divergence of the
pseudotensor to get \be\lb{82a} \Delta P_{\mu }=- \frac{1}{32\pi
G}\int d^4x h_{\alpha\beta,\mu }\psi^{\alpha\beta,\nu }_{,\nu}. \ee
One can show that the lowest order giving non-zero contribution is
the second (or the first post-linear order). Using the Einstein
equations in quasilinear form, as given in the second section, we
perform transformations similarly to the  electromagnetic case
introducing the polarization tensors for gravitational waves. The
final expression for the loss of the four-momentum on gravitational
radiation with the polarization
 $\lambda$ reads:
\be\lb{84} \Delta P_{gr}^{(\lambda)\mu }=\frac{G}{\pi ^2}\int
d^4kk^{\mu } \theta (k_0) \delta(k^2)\mid \tau ^{(\lambda )}(k)\mid
^2\ , \ee where $\tau^{(\lambda )}=e_{\m }^{(\lambda )}\tau ^{\m },\
\tau ^{\m }= T^{\m } + S^{\m }$,
\begin{align} \lb{85}-16\pi GS_{\m }= &\, h_{\mu }^{\alpha ,\beta }h_{\nu \beta ,\alpha }-
h_{\mu }^{\alpha ,\beta }h_{\nu \alpha ,\beta }- (1/2)h_{,\mu
}^{\alpha \beta }h_{\alpha \beta ,\nu }+\nn \\& + h^{\alpha \beta
}(h_{\mu \alpha ,\nu ,\beta }+h_{\nu \alpha ,\mu ,\beta }-
h_{\alpha \beta ,\mu ,\nu }-h_{\mu \nu ,\alpha , \beta })-\nn
\\& -(1/2)h_{,\alpha }^{,\alpha }h_{\m }+(1/2)\eta _{\m
}(2h^{\alpha \beta } h_{\alpha \beta ,\lambda }^{,\lambda
}-h_{\alpha \beta ,\lambda } h^{\alpha \lambda ,\beta
}+(3/2)h_{\alpha \beta ,\lambda }h^{\alpha \beta , \lambda })\ ,
\end{align} and it is assumed that all contractions over indices
in (\ref{84}) and (\ref{85}) are performed with Minkowski metric
$\eta _{\m }$.

It is convenient to choose as two independent polarization vectors
the quantities
\be\lb{86} 
{ e^{\times }_{\m }=(1/\sqrt{2})(e_{\mu }^{\theta }e_{\nu }^{\phi
}+ e_{\nu }^{\theta }e_{\mu }^{\phi }), e_{\m
}^{+}=(1/\sqrt{2})(e_{\mu }^{\theta }e_{\nu }^{\theta }- e_{\mu
}^{\phi }e_{\nu }^{\phi }), } \ee \be\lb{87} e_{\m }^{\lambda
}e^{\lambda \mu \nu }=1,\ e_{\mu }^{\lambda \mu }=0,\ e_{\m
}^{\lambda }=e_{\nu \mu }^{\lambda },\ k^{\mu }e_{\m }^{\lambda
}=0,\ \lambda = \times ,+\ . \ee The subsequent calculations are
essentially similar (though more lengthy) as for the scalar and
electromagnetic radiation, so we give the final result.  The
amplitudes $T^{\m }$ and $S^{\m }$ in an arbitrary Lorentz frame
read
$$T^{\m }(k)=T_{1}^{\m }(k)+T_2^{\m }(k)\ , $$
where \begin{align}\lb{88}
 T_1^{\m }(k)=& 4Ge^{ik\Delta }\biggl\{
\biggl[ - \biggl( 1-\frac{(m_1m_2)^2}{2I^2} \biggr)
\frac{(p_1p_2)}{I}\frac{m_2}{m_1}\frac{z_2}{z_1}+\biggl(1+\frac{(m_1m_2)^2}{2I^2}\biggr)
\frac{m_2^2}{I}\biggr] K_0(z_1)p_1^{\mu } p_1^{\nu }-\nn \\
& -\biggl(1+\frac{(m_1m_2)^2}{2I^2}\biggr)
\frac{(p_1p_2)}{I}K_0(z_1)(p_1^{\mu } p_2^{\nu }+p_1^{\nu
}p_2^{\mu })+i(k\Delta )\biggl(1+\frac{(m_1m_2)^2}{2I^2}\biggr)
\frac{m_2^2}{I}\frac{K_1(z_1)}
{z_1}p_1^{\mu }p_1^{\nu }-\nn \\
& -i(kp_1)\biggl(1+\frac{(m_1m_2)^2}{2I^2}\biggr)
\frac{m_2^2}{I}\frac{K_1(z_1)}{z_1} (p_1^{\mu }\Delta ^{\nu
}+p_1^{\nu }\Delta ^{\mu })\biggr\}\ , \end{align}
$$T_2^{\m }(k)=e^{+ik\Delta }T_1^{\mu \nu ^{*}}(1\leftrightarrow 2)\ ,$$
\begin{align} \lb{89} &S^{\m }(k)=4GIe^{ik\Delta }\!\! \int\limits_0^1dxe^{-ix(k\Delta
)}\Biggl\{\! \biggl( \!1+\frac{(m_1m_2)^2}{2I^2}\biggr)
z(x)K_1(z(x))\frac{\Delta ^{\mu }\Delta ^{\nu }}{\Delta ^2}-
\biggl(1+\frac{(m_1m_2)^2}{2I^2}\biggr)
\frac{m_2^2}{I^2}K_0(z(x))p_1^{\mu } p_1^{\nu }+\nn \\
& \,\,+\frac{z_2^2}{m_1^2}\biggl[ \! \biggl(
1-\frac{m_1m_2(p_1p_2)}{I^2}\biggl(\frac{
(p_1p_2)}{m_1m_2}z_2x+z_1(1-x)\biggr) \!\biggr)
^{\!2}\!-\!\frac{(m_1m_2)^4}{2I^4} \biggl(
\frac{(p_1p_2)}{m_1m_2}z_2x+z_1(1-x)\biggr)^{\!2} \biggr]
\frac{K_1(z(x))}{z(x)}p_1^{\mu }p_1^{\nu }+\nn \\&\,\,
+iz_2\frac{m_2}{I}\biggl[ \frac{(p_1p_2)}{m_1m_2}- \biggl(
1+\frac{(m_1m_2)^2}{2I^2}\biggr) \biggl(
\frac{(p_1p_2)}{m_1m_2}z_2x+z_1(1-x)\biggr) \biggr]
K_0(z(x))\biggl( p_1^{\mu }\frac{\Delta ^{\nu }}{\sqrt{-\Delta
^2}}+ p_1^{\nu }\frac{\Delta ^{\mu }}{\sqrt{-\Delta ^2}} \biggr)
\Biggr\}\ . \end{align}

Note that in the electromagnetic case the local and non-local
currents are separately gauge invariant, while in the gravitational
case  only the sum   $T_1^{\m }+T_2^{\m }+S^{\m }$ is independent on
the gauge choice. This allows to change  contribution from separate
terms choosing suitable gauge. In particular, in the gauge
(\ref{86}) the contribution from $T_1^{\m }$ is zero. The subsequent
calculation will be performed in the rest  frame of the second mass
$p_2^{\mu }=(m_2,0)$.

In the non-relativistic limit $(v\ll 1)~ z(x)\approx z_1\approx
z_2\approx \omega \rho /v$~ , so the integral (\ref{89}) can be
easily computed. The contributions from  (\ref{88}) and (\ref{89})
turn out to be of the same order, and taking into account
(\ref{84}), one finds: \begin{align} \lb{90}
\frac{d^2E_{gr}^{+}}{d\omega d\Omega }=\frac{G^3(m_1m_2)^2}{\pi ^2
\rho ^2} a^2 & \left\{ 4\sin ^2 \theta \cos ^2 \theta \cos ^2\phi
[K_1(a)+K_0(a)]^2+ \right. \nn \\ &+[\sin ^2 \theta K_0 (a) +(\sin
^2 \theta +\sin ^2\phi -\cos^2\theta \cos ^2\phi ) aK_1(a)]^2\} \
,
\end{align}
\begin{align} \lb{91} \frac{d^2E_{gr}^{\times }}{d\omega d\Omega }=4\frac{G^3(m_1m_2)^2}{\pi ^2\rho ^2}
a^2\{ \cos ^2\theta \cos ^2 \phi a^2 K_1^2(a)+\sin ^2\theta \sin
^2\phi [K_1(a)+aK_0(a)]^2\}\ , \end{align} and after the
integration \be\lb{92} \Delta E_{gr}^{+}=\frac{4327\pi }{3840
}\frac{G^3(m_1m_2)^2v}{\rho ^3}\ ,\quad \Delta E_{gr }^{\times
}=\frac{343\pi }{256}\frac{G^3(m_1m_2)^2v}{\rho ^3}\ . \ee The
Eqs. (\ref{90}) and (\ref{91}) coincide with those given in
\cite{KT3}.

An ultrarelativistic case is considered similarly to the previous
sections. For $\omega \ll \rho ^{-1}$ we obtain an expression
coinciding with the result of application of the low frequency
theorems \cite{We}
\be\lb{93} \frac{dE_{gr }^{+}}{d\omega }=\frac{32G^3(m_1m_2)^2}{3\pi
}\biggl( \frac{\gamma } {\rho }\biggr) ^2\ , \ee \be\lb{94}
\frac{dE_{gr}^{\times }}{d\omega }=\frac{64G^3(m_1m_2)^2}{\pi
}\biggl( \frac{ \gamma }{\rho }\biggr) ^2 \biggl( \ln 2\gamma
-\frac{1}{2} \biggr)\ . \ee Note different dependence of (\ref{93})
and  (\ref{94}) on the energy.

For the frequencies $\omega \geqslant \rho ^{-1}$ the leading in
$\gamma $ approximation the spectral distribution of the
gravitational bremsstrahlung summed up over polarizations is given
by \be\lb{95} \frac{dE_{gr}}{d\omega }=\frac{16G^3(m_1m_2)^2}{\pi
}\omega ^2 \int \limits_0^{\infty } \int \limits_0^{\infty }d\xi
d\eta e^{-\frac{2\omega \rho }{\gamma }\sqrt{1+\xi ^2}
\sqrt{1+\eta ^2}}\frac{1+8\xi ^2+8\xi ^4}{(1+\xi ^2)^{3/2}(1+\eta
^2)^{1/2} } \ln \frac{1+\xi ^2 +\eta ^2}{\eta ^2 }\ . \ee In
contrast to the previous cases, the spectral curve is monotonous
function of the frequency, and for relatively small frequencies
$\rho ^{-1} \le \omega \ll \gamma /\rho $ the spectrum falls off
logarithmically \be\lb{96} \frac{dE_{gr}}{d\omega
}=\frac{64G^3(m_1m_2)^2}{\pi }\biggl( \frac{\gamma } {\rho }
\biggr) ^2 \ln \frac{2\gamma }{e^C\omega \rho }\ , \ee while for
$\omega \gg \gamma /\rho $ -- the fall off is exponential
\be\lb{97} \frac{dE_{gr}}{d\omega }=4G^3(m_1m_2)^2\biggl(
\frac{\gamma } {\rho } \biggr) ^2 \biggl( \frac{\omega \rho
}{\gamma }\biggr)
 \ln \frac{4e^C\omega \rho }{\gamma }e^{-\frac{2\omega \rho }{\gamma }}\ .
\ee
 Note that for  $\omega =\rho ^{-1}$ (\ref{96}) with
logarithmic accuracy coincides with (\ref{94}), while (\ref{96}) and
(\ref{97}) by the order of magnitude are compatible at $\omega
\gamma /\rho $. So the Eqs. (\ref{94}), (\ref{96}) and (\ref{97})
together covers the whole frequency spectrum. The total radiated
energy is obtained integrating (\ref{95}) over frequencies
$$ \Delta E_{gr}=\Lambda _{gr} G^3(m_1m_2)^2(\gamma /\rho )^3\ ,$$
\be\lb{98} \Lambda _{gr}=\pi +35\tilde G/2-211/12 \approx 29\ . \ee
The result (\ref{98}) by the order of magnitude coincides the the
results of refs. \cite{KT3, KT4, Pe1, DE}, but differ from
\cite{MaNu, Sm2} by the absence of the factor $\ln (2\gamma )$.
 The frequency
distributions of scalar, electromagnetic and gravitational radiation
under ultrarelativistic gravitational scattering are shown in Fig.
1.
\begin{figure}
\begin{center}
\includegraphics[angle=0,width=10cm]{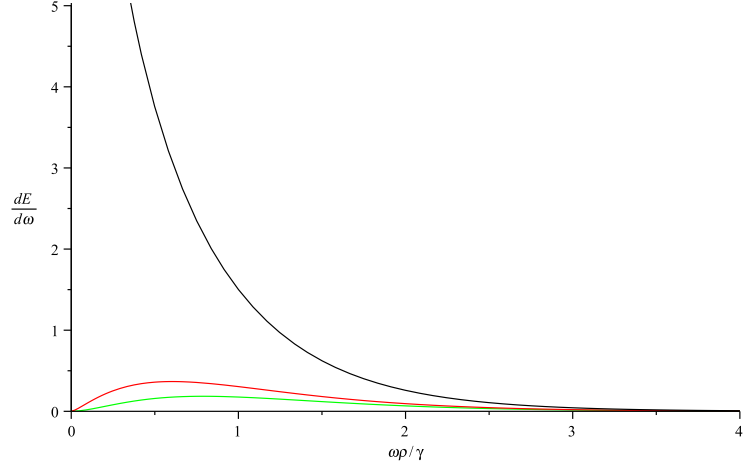}
\caption{The spectral distribution of scalar (green),
electromagnetic (red)  and gravitational (black) radiation under
gravitational scattering for $\gamma=1000$.}
  \label{freq_dist}
\end{center}
\end{figure}

Consider now for comparison gravitational radiation under collision
mediated by non-gravitational forces. Let both particles be charged
(with $e_1$ and $e_2$ correspondingly) with large charge to mass
ratio, so their gravitational interaction can be neglected. Then in
the lowest in $G$ approximation $g_{\m }=\eta _{\m },\ S_{\m } =0$,
and $\tau _{\m } =T_{\m }+ _FT_{\m }$, where $ _FT_{\m }$ is the
energy-momentum tensor of the electromagnetic field. Calculations
shows   that with the same accuracy
\begin{align}
T_1^{\m }(k)=\frac{2e_1e_2m_2^2}{I^3}e^{ik\Delta }\biggl\{&
-\biggl( (p_1p_2) +m_1m_2\frac{z_2}{z_1} \biggr) K_0(z_1)p_1^{\mu
}p_1^{\nu } +m_1^2K_0(z_1)(p_1^{\mu }p_2^{\nu }+p_1^{\nu }p_2^{\mu
})-\nn \\ &-i(k\Delta )(p_1p_2) \frac{K_1(z_1)}{z_1}p_1^{\mu
}p_1^{\nu } +i(kp_1)(p_1p_2)\frac{K_1(z_1)}{z_1}(p_1^{\mu }\Delta
^{\nu }+p_1^{\nu } \Delta ^{\mu }) \biggr\}\ ,
\end{align}$$
T_2^{\m }(k) =e^{ik\Delta }T_1^{\mu \nu ^{*}}(1 \leftrightarrow
2)\ , \nn$$
\begin{align}
_FT^{\m } (k)&= 2 e_1 e_2  e^{ik\Delta }\int _0^1 dx e^{-ix(k\Delta
)} \biggl\{ \frac{(p_1p_2)}{I}\biggl[ z(x)K_1(z(x)) \frac{\Delta
^{\mu } \Delta ^{\nu }}{-\Delta ^2}
+\frac{(m_1m_2)^2}{I^2}K_0(z(x))\frac{p_1^{\mu }p_1^{\nu }}{m_1^2}
\biggr] -\nn \\ &-\frac{(m_1m_2)^2}{I^2}\biggl(
\frac{(p_1p_2)}{m_1m_2}z_2x +z_1(1-x) \biggr) \biggl[
\frac{(p_1p_2)}{I}\biggl( \frac{(p_1p_2)}{m_1m_2}z_2x +z_1(1-x)
\biggr) -\frac{I}{m_1m_2}z_2\biggr] \frac{K_1}{z(x)} \frac{p_1^{\mu
}p_2^{\nu }}{m_1^2}+\nn \\& +i\frac{m_1m_2}{I}\biggl[ \frac{
(p_1p_2)}{I} \biggl( \frac{(p_1p_2)}{m_1m_2}z_2x +z_1(1-x) \biggr)
-\frac{I}{2m_1m_2}z_2\biggr]K_0 \biggl( \frac{\Delta ^{\mu
}}{\sqrt{-\Delta ^2}}\frac{p_1^{\nu }}{m_1} +\frac{\Delta ^{\nu
}}{\sqrt{-\Delta ^2}}\frac{p_1^{\mu }}{m_1} \biggr) \biggr\}\ ,
\end{align}
and in the chosen gauge the contribution form  $T_2^{\m }$ is
zero. For $\gamma \gg 1$ in the frequency region $\omega \ll \rho
^{-1}$ one obtains the results (\ref{96},\ref{97}), in which the
gravitational deflection angle should be replaced by the
electromagnetic one $4Gm_1m_2\gamma \rightarrow 2e_1e_2$.

For  $\omega \geqslant \rho ^{-1}$ the non-locality of the source
due to presence of the term $ _FT^{\m }$ leads to destructive
interference. Like in the above cases of the gravitational
interaction the non-local source  $_FT^{\m }$ gives rise to two
impulses of different duration, one of which comes to the
observation point in the counterphase with the one due to  $T^{\m
}$. But contrary to the case of gravitational interaction, the
contribution of   $T^{\m }$ is canceled only partially. As a
result, the spectral-angular distribution for  $\omega \geqslant
\rho ^{-1}$ and $\gamma \gg 1$will be \be\lb{99}
\frac{d^2E_{gr}}{d\omega d\cos \theta }=\frac{2G(e_1e_2)^2}{\pi
}\omega ^2 K_1^2(\omega \rho \delta )\sin ^2 \theta \biggl(
1-\frac{\sin ^2\theta }{ 2\gamma ^2\delta ^2}\biggr)\ . \ee In the
leading in  $\gamma $ order we obtain \be\lb{100}
\frac{dE_{gr}}{d\omega }=\frac{4G(e_1e_2)^2}{\pi \rho ^2}z\int
_z^{\infty }dx
 K_1^2(x) \biggl( \frac{4z}{x}-\frac{2z^2}{x^2}-3+\frac{x}{z}\biggr)
\ee which has the following asymptotic behavior for $\rho ^{-1}\le
\omega \ll \gamma ^2/\rho $ \be\lb{101} \frac{dE_{gr}}{d\omega
}=\frac{4G(e_1e_2)^2}{\pi \rho ^2 }\ln \frac{4\gamma ^2}{ e^C\omega
\rho }\ , \ee for $\omega \gg \gamma ^2/\rho $ \be\lb{102}
\frac{dE_{gr}}{d\omega }=\frac{G(e_1e_2)^2}{\rho ^2}\biggl(
\frac{\gamma ^2} {\omega \rho }\biggr) e^{-\frac{\omega \rho
}{\gamma ^2}}\ . \ee

The total energy loss is \be\lb{103} \Delta E_{gr}=(\pi /4)\gamma ^2
G(e_1e_2)^2/\rho ^3\ . \ee The result (\ref{103}) coincides with
that of \cite{Pe3}.

The spectral distribution of gravitational radiation in two
considered cases has the following distinctive features. For $\omega
\ll \rho ^{-1}$ it weakly depends on frequency and for the fixed
deviation angle depends on the energy as  $\gamma ^2\ln 2\gamma $.
For the frequencies $\rho ^{-1} \le \omega \ll \omega _{cr}$ the
spectrum falls off logarithmically, while for $\omega \gg \omega
_{cr} $ exponentially. But if for the electromagnetic interaction
$\omega _{cr}^{em}=2\gamma ^2/\rho $, for gravitational interaction
 $\omega _{cr}^{gr} =\gamma /\rho $. Also, for the same scattering
angle the total radiated energy for electromagnetic interaction the
radiative loss is  $\gamma $ larger that for gravitational
interaction.

\section{Low frequency limit}
For $\omega \rightarrow 0$ the spectral distribution does not
depend on frequency and one could hope to get a correct estimate
for the energy loss under collision multiplying the Eq. (\ref{68})
or (\ref{93}) and (\ref{94}) on a suitable frequency cutoff. For
radiation of the point particle in the flat space (in the case of
non-gravitational interaction) the cutoff frequency in the
classical spectrum is estimated kinematically as an inverse time
of the formation of radiation in the given direction and it is
given by $\omega _{cr}^{em} \sim \gamma ^2/\rho $. In the
gravitational case a similar estimate  is $\omega _{cr}^{gr} \sim
\gamma /\rho$ which is confirmed by an accurate calculation. Now,
it can be easily seen that the low frequency approximation gives a
correct estimate of the total radiated energy in the
electromagnetic case, but  gives an wrong factor  $\ln 2\gamma $
in the gravitational case. The reason of this discrepancy lies in
the fact that the fall-off in the spectral distribution in the
gravitational case corresponds not to the frequency $\omega
_{cr}$, as it is assumed in the low-frequency approach, but to
$\omega \sim \rho ^{-1}$ (see (\ref{96}) and (\ref{101})).
Logarithmic fall-off in the high frequency region $\omega
\geqslant \rho ^{-1}$ cancels an extra logarithmic factor. This
explains the difference between our result (\ref{98}) and that of
\cite{Sm1,Sm2}.
\section{ Method of virtual gravitons}
The spectral density of the wave packet of equivalent gravitons
imitating the gravitational field of the ultrarelativistic particle
is given by \be \label{104} I_{gr} (\omega ,\rho )=G(m_2/\pi \rho
)^2(\omega \rho /\gamma )^2K_2^2 (\omega \rho /\gamma ) \ee (this
results differs from  that of the ref. \cite{MaNu} by a numerical
factor). The spectrum diverges for $\omega \rightarrow 0$. This
means that it can be applied only for sufficiently high frequencies.
Applying this spectrum to compute bremsstrahlung (electromagnetic or
gravitational) under scattering of the fast particle on the fixed
center one has to introduce the frequency cutoff. The results differ
from those obtained in this paper by a factor $\ln 2\gamma $. Thus,
contrary to the electromagnetic case, where  method of virtual
quanta gives the correct answer in the ultrarelativistic limit, in
the gravitational case this method fails. The reason is that the
spectrum of virtual gravitons describes correctly the frequency
range $\omega \gg \gamma /\rho$, which, as we have seen, is
negligible in the total radiation due to non-locality of the
effective radiation sources. Indeed, let us consider radiation in
the forward direction. For $\theta =0$ the integral over the Feynman
parameter can be computed exactly and we obtain \be
\label{105}\left.\frac{d^2 E_{gr}}{d\omega d\Omega }\right|_{\theta
=0}=\frac{G^3(m_1m_2)^2} {\pi ^2\gamma ^2} \omega ^2 \biggl[
K_2(z_1)-\biggl( \frac{z_2}{z_1}\biggr) ^2K_1(z_2) \biggr] ^2\ . \ee
At the same time the equivalent gravitons approach gives \be\lb{106}
\left.\frac{d^2E_{gr}}{d\omega d\Omega } \right|_{\theta
=0}=\frac{G^3(m_1m_2)^2} {\pi ^2 \gamma ^2}\omega ^2 K_2^2(z_1)\ .
\ee One can see that the expressions (\ref{105}) and (\ref{106}) are
compatible only for  $\omega \gg \gamma /\rho $.

\section{Conclusions}

We have presented Lorentz-covariant perturbation approach in General
Relativity using the momentum space formulation similar to quantum
field theory perturbation theory. The method consists in solving
particles equations of motion and the field equations iteratively in
terms of the gravitational coupling constant. Gravitational
radiation arises in the second order approximation. In terms of the
flat space metric the source of the D'Alembert equation for the
second order metric perturbation is non-local and contains the
contribution from gravitational stresses computed in the first
order. This non-locality results in $\gamma$ times lower frequency
cutoff as compared to  the case of non-gravitational interaction.
For this reason the method of virtual gravitons is not applicable
for gravitational scattering of ultrarelativistic particles. The
total energy loss in the rest frame of one of the particles is
proportional to the third order of $\gamma$. Radiation from two
colliding bodies looks as a collective effect, contributions form
each of them can not be separated in a gauge independent way.

\end{document}